\documentclass[fleqn,usenatbib]{mnras}

\usepackage{newtxtext,newtxmath}

\usepackage[T1]{fontenc}
\usepackage{ae,aecompl}

\usepackage{graphicx}	
\usepackage{amsmath}	
\usepackage{amssymb}	
\usepackage{bm}

\title[Disc formation]{Misaligned accretion disc formation via  Kozai--Lidov oscillations}

\author[A. Franchini et al.]{
Alessia Franchini$^{1}$\thanks{E-mail: alessia.franchini@unlv.edu},
Rebecca G. Martin$^{1},$ and
Stephen H. Lubow$^{2}$
\\
$^{1}$Department of Physics and Astronomy, University of Nevada, 4505 South Maryland Parkway, Las Vegas, NV 89154, USA\\
$^{2}$Space Telescope Science Institute, 3700 San Martin Drive, Baltimore, MD 21218, USA\\
}

\date{Accepted XXX. Received YYY; in original form ZZZ}

\pubyear{2018}

\begin{document}
\label{firstpage}
\pagerange{\pageref{firstpage}--\pageref{lastpage}}
\maketitle

\begin{abstract}
We investigate the formation and evolution of misaligned accretion discs around the secondary component of a binary through mass transfer driven by Kozai-Lidov oscillations of the circumprimary disc's eccentricity and inclination. We perform SPH simulations to study the amount of mass transferred to the secondary star as a function of both the disc and binary parameters. For the range of parameters we explore, we find that increasing the disc aspect ratio, viscosity parameter and initial inclination as well as decreasing the binary mass ratio leads to larger amount of mass transfer, up to a maximum of about ten per cent of the initial mass of the primary disc.  The  circumsecondary disc forms with a high eccentricity and a high inclination and is also able to undergo KL oscillations. The circumsecondary disc oscillations have a shorter period than those in the disc around the primary. We find that some of the material that escapes the Roche-lobe of the two components forms a 
misaligned circumbinary accretion disc. 
This study has implications for disc evolution in young binary star systems.
\end{abstract}




\section{Introduction}
\label{intro}
The recent improvements in observational technology have enabled the  discovery of thousands of exoplanets over the last 30 years.
Roughly 40\%-50\% of them are expected to be hosted in binary star systems \citep{Horch2014}.
It is therefore important to understand planet formation and evolution in binary systems in order to be able to explain exoplanet properties. Furthermore, it is crucial to study the birthplace of planets, the  protoplanetary discs, to answer many questions about exoplanetary systems.


In a young binary star system, the accretion disc around each star may be misaligned with respect to the plane of the binary.
There are a few observed misalignments in wide binary systems with separations greater than $40\,$au  \citep{Jensen2004,Roccatagliata2011}.
HST and recent ALMA observations showed a misalignment between the two accretion discs in the young system HK Tau \citep{Stapelfeldt1998,Jensen2014}.
This suggests that during the formation process these systems might be subjected to small scale perturbations, such as turbulence \citep{Offner2010,Bate2010,Bate2012}, that can result in the misalignment between the gas orbiting the two components of the binary. Numerical simulations have also shown that accretion of material onto forming stars occurs in different directions over time. Therefore, misaligned accretion discs are likely to be typical rather than rare cases \citep{Bate2009,Bate2012,Bate2018}.

To date, hundreds of planets have been also observed to lie in eccentric orbits with eccentricities up to about $0.9$ \citep{Tamuz2008,O'Toole2009,Moutou2009}. 
A planet embedded in a gaseous disc will undergo damping of eccentricity and inclination \citep{Xiang2013} unless it is subject to planet-planet scattering or secular perturbations \citep{Ford2008,Lubow2016}. In this paper we investigate the effects of secular perturbations, specifically the Kozai-Lidov (KL) mechanism \citep{Kozai1962,Lidov1962}. 
This is a secular effect that likely explains the eccentricities and inclinations observed in exoplanets \citep{Wu2003}.
This process periodically exchanges the eccentricity with the inclination of a misaligned test particle around one component of a binary system, if its initial inclination is above a certain critical angle.

For the first time, \cite{Martin2014} found that the KL mechanism can operate also in a fluid accretion disc. The disc remains relatively flat, i.e. not strongly warped, and the eccentricity is roughly constant in radius. The disc then responds in a global fashion to the KL oscillations.  
\cite{Fu2015a} explored further the parameter space and found that the oscillations occur over a fairly broad range of disc and binary parameters.
The oscillations are damped on a timescale of a few tens of binary orbits due to the effect of the disc viscosity \citep[e.g.,][]{King2013}.
The alignment timescale may be longer than the lifetime of the disc, meaning that some discs are likely to be dispersed before they have a chance to align with the orbital plane. If a planet remains at or below the critical KL angle after the disc has been dispersed, the planet will remain stable against KL oscillations \citep{Lubow2016,Martin2016}. 

Disc self-gravity can suppress KL disc oscillations \citep{Batygin2012,Fu2015b,Fu2017}. Self-gravity likely plays an important role
during the earliest phases of star formation, prior to the T Tauri phase \citep{Kratter2016}. As the disc accretes without being resupplied by infalling gas, its level of self-gravity is reduced
and KL oscillations can commence. We report here on the evolution
of a misaligned disc at this stage, idealised by neglecting the disc self-gravity. 

In this paper, we explore for the first time a process by which circumprimary disc
material is transferred to a secondary star that does not initially have a disc. We consider 
an initially highly misaligned circular circumprimary disc. The disc undergoes an increase of its radial extent due to the substantial eccentricity it acquires during KL oscillations. 
If this circumprimary disc extends beyond the Roche lobe of the primary star, it can then transfer mass to the companion star.
This Roche lobe overflow can lead to the formation of an accretion disc around the secondary star.
The investigation of the mass transfer mechanism is important for understanding the evolution of misaligned accretion discs in binaries. Furthermore,  this process may affect the conditions for planet formation.

The outline of the paper is as follows.
In Section~\ref{sec:eccKL} we consider inclined test particle orbits around one component of binary as a guide to some properties of a disc. In Section \ref{sec:sim} we present the results of SPH simulations for the formation of the circumsecondary disc by mass transfer from the primary disc that is undergoing KL oscillations. In Section \ref{sec:circbin} we discuss the formation of a circumbinary disc as a result of the KL oscillation of the circumprimary disc (and possibly the circumsecondary disc). Finally, we draw our conclusions in Section \ref{sec:concl}.

\section{Test particle orbits}
\label{sec:eccKL}

\begin{figure}
\centering
\includegraphics[width=\columnwidth]{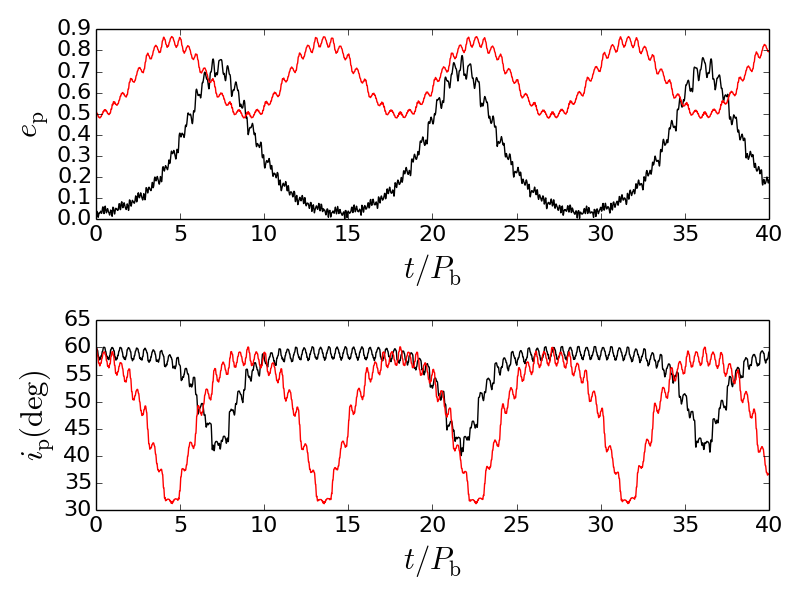}
\caption{Eccentricity and inclination evolution of a test particle around one of the component of an equal mass circular orbit binary. The particle is initially at apastron at distance of $d = 0.2 a$ from the  secondary component. The black line refers to an initially circular particle orbit, $e_{\rm p0}=0$, while the red line represents the case with initial non-zero eccentricity, $e_{\rm p0} =0.5$ and initial argument of periapsis $\omega_0=0^\circ$.}
\label{fig:testp}
\end{figure}

We consider inclined test particle orbits around one component of a  circular orbit binary with semi-major axis $a$.
An initially circular test particle orbit undergoes KL oscillations if the initial inclination $i_{\rm p0}$ is in the range
\begin{equation}
\cos^2 i_{\rm p0} < \cos^2i_{\rm cr} = 3/5 \,,
\label{icr}
\end{equation}
which corresponds to the inclination range $39^{\circ} \leq i_{\rm p0} \leq 141^{\circ}$ \citep{Kozai1962,Lidov1962}.
From this condition it follows that there is an upper limit on the eccentricity that can be achieved during these oscillations \citep{Innanen1997}.
Over long timescales, a particle's orbital energy and therefore its semi-major axis $a_{\rm p}$ are conserved because the secular potential is static.
In addition, the component of the angular momentum perpendicular to the binary orbital plane is conserved. These two conditions imply
\begin{equation}
\sqrt{1-e_{\rm p}^2}\,\cos i_{\rm p} \approx {\rm const}\,,
\label{eq:angmomconst}
\end{equation}
where $e_{\rm p}$ and $i_{\rm p}$ are the eccentricity and the inclination of a test particle orbiting one of the binary stars.
The eccentricity of the particle increases from $e=0$ up to
\begin{equation}
e_{\rm max} = \sqrt{1-\frac{5}{3}\cos^2i_{\rm p0}}\,,
\label{eq:emax}
\end{equation}
while its inclination decreases from $i_{\rm p0}$ to $i_{\rm cr}$. 





Test particle orbits provide some indication of the behaviour of fluid
discs undergoing KL oscillations in 
hydrodynamical simulations \citep{Martin2014}.    The disc KL eccentricity approximately agrees that of particles. However,  dissipation limits the disc eccentricity, especially at high eccentricity.
Particle orbits also provide an estimate of the disc tilt oscillation frequency
when suitably averaged in radius. The misaligned circumprimary disc is  assumed to be initially circular and therefore its eccentricity oscillates from $0$ to roughly $e_{\rm max}$ for a particle given by Equation (\ref{eq:emax}), provided that $e_{\rm max}$ is mild. 

As we will see in the next section, the circumsecondary disc that forms through mass transfer has an initial non-zero eccentricity.
In order to understand the disc
properties, we need to take into account the initial non-zero eccentricity of the disc.
To obtain an estimate of the KL oscillation period 
in case of an initially eccentric accretion disc, we can carry out test particle calculations.  

We consider an equal mass binary and so either star could be regarded as primary.
Since we are interested 
in circumsecondary disc formation, we regard the particle to be orbiting the secondary star.
We define a reference direction to be along a line  from the primary star to the secondary star
at the initial time. We determine particle orbits that start with inclination $i_0=60^\circ$, longitude of ascending node $0^\circ$,
true anomaly of $180^\circ$ (apastron), distance from secondary $d=0.2 a$, and
argument of periapsis $\omega_0=0^\circ$.
Figure \ref{fig:testp} shows the eccentricity and inclination evolution for such a test particle on an initially circular ($e_{\rm p0} =0$) orbit and initially eccentric  ($e_{\rm p0} =0.5$) orbit.  

Comparing the black (initially circular) and red (initially eccentric) curves in Figure \ref{fig:testp}, we see that the period of the KL oscillations  for the initially eccentric orbit case is shorter than the initially circular case by roughly a factor 2,  for same initial apastron radius. The KL oscillation period depends on tidal torques that
vary with radius. 

We have verified that the results are insensitive to the initial value of the longitude
of the ascending node, as is expected by secular theory \citep{Fabrycky2007}.  We also find that the  KL oscillation periods 
for initially eccentric particle orbits for $e_{\rm p0}=0.5$ are shorter for other initial angles of periapsis,
$\omega_0=90^\circ$, $180^\circ$, and $270^\circ$, with initial distance $d=0.2a$ at apastron.

Another effect associated with starting at nonzero particle eccentricity is that the minimum tilt angle $i_{\rm min}$ generally differs from the $i_{\rm cr}=39.2^\circ$ value
for the initially circular orbit case, as seen in Figure~\ref{fig:testp}. As seen in upper panel of Figure 2 in \cite{Fabrycky2007}, $i_{\rm min}$ is less than
$i_{\rm cr}$ for $ 45^\circ > \omega_0 > 0$ and $i_{\rm min}$ is greater than
$i_{\rm cr}$ for $ 90^\circ > \omega_0 > 45^\circ$. 

In Section \ref{sec:sim} we show through SPH simulations that the  circumsecondary disc forms with non-zero eccentricity and undergoes KL oscillations with a period that is shorter by roughly a factor 2 with respect to the KL period of the (initially circular) circumprimary disc, in agreement with our test particle calculations.

\begin{figure}
\centering
\includegraphics[width=0.9\columnwidth]{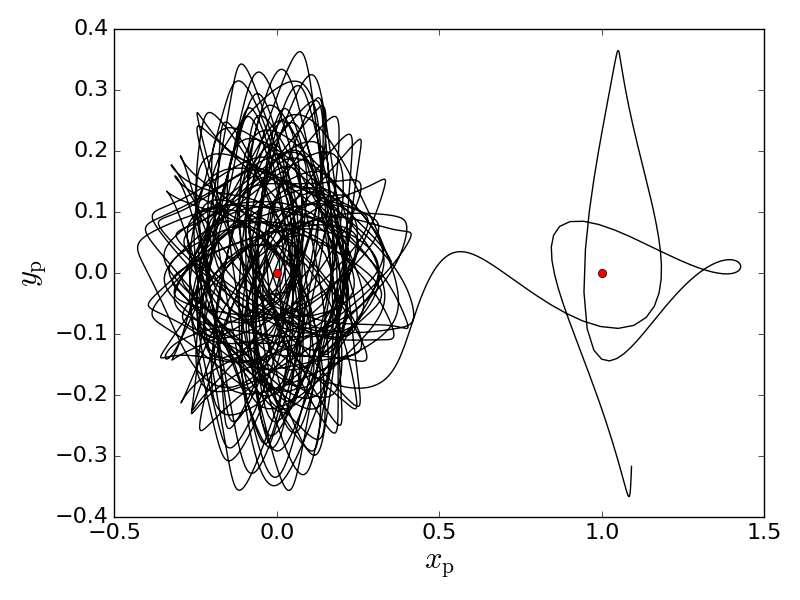}
\caption{Test particle trajectory  in the $x_{\rm p}-z_{\rm p}$ plane that begins around the primary star at  distance $0.36$a from the star with initial tilt of $i_{\rm p0}= 60^\circ$.
The particle gains eccentricity and eventually is transferred to the secondary star. 
The particle trajectory is identified by the black lines and the binary component by the two red dots. }
\label{fig:ballistic}
\end{figure}

\begin{figure}
\centering
 	\includegraphics[width=\columnwidth]{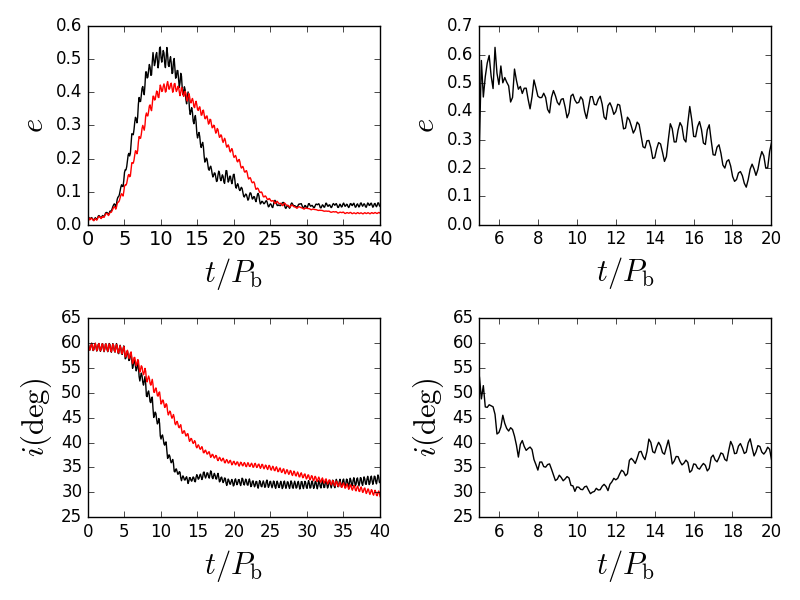}
\includegraphics[width=0.8\columnwidth]{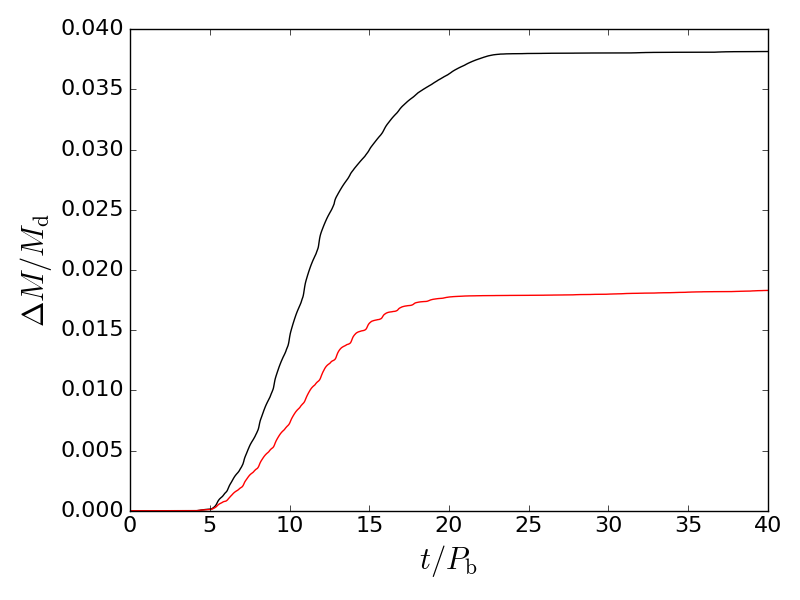}
    \caption{Top four panels: Eccentricity (upper panels) and inclination (lower panels) evolution of the circumprimary (left) and circumsecondary (right) disc averaged over the whole disc. 
    Bottom panel: Accreted mass within a distance of $R_{\rm acc}=0.025a$ from the secondary star scaled by the initial circumprimary disc mass $\Delta M / M_{\rm d}$ as a function of time in units of $P_{\rm b}$. 
    The black lines refer to the  disc sounds speed described in Equation (\ref{eq:farris}) with $q=0.5$  (run2),  while the red lines are for a globally isothermal simulation ($q=0$, run1).}
    \label{fig:macc_eos_14_1}
\end{figure}

\begin{figure}
\centering
   \includegraphics[width=0.8\columnwidth]{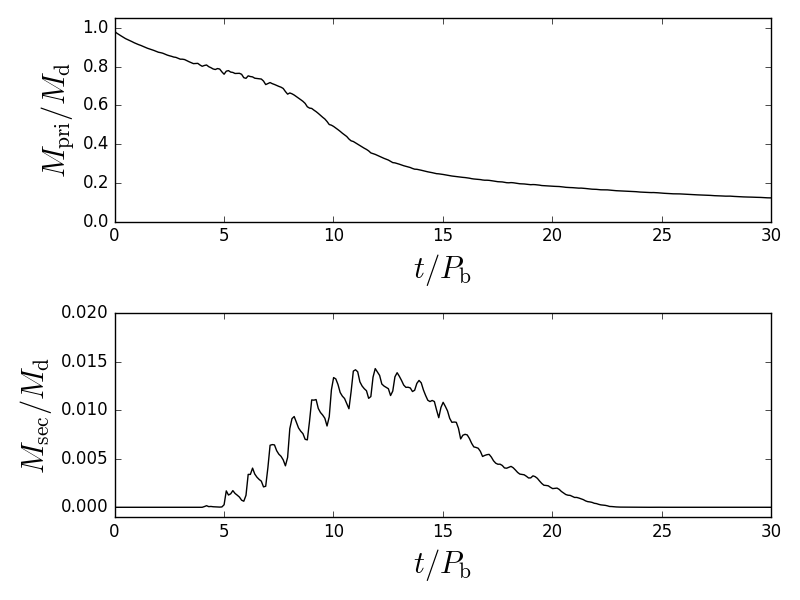}
   \caption{Mass contained in the circumprimary (upper panel) and circumsecondary (lower panel) discs outside of $R_{\rm acc} =0.025a$ from the central star in units of the initial circumprimary disc mass $M_{\rm d}=0.001$ as a function of time in units of $P_{\rm b}$. The disc sounds speed is described in Equation (\ref{eq:farris}) with $q=0.5$  (run2).}
   \label{fig:mdisc_eos14_1}
\end{figure}

\section{Circumsecondary disc formation}
\label{sec:sim} 

We show here that a misaligned accretion disc around a primary star that undergoes KL oscillations of eccentricity and inclination can become eccentric enough to overflow its Roche lobe and transfer mass onto the secondary companion. 
We define a Cartesian coordinate system $(x_{\rm p}, y_{\rm p}, z_{\rm p})$ that is  centered on the primary and corotates with the binary. The $x_{\rm p}-$axis is along the line that joins
the two stars. The $z_{\rm p}$-axis is parallel to the binary angular momentum.
Figure \ref{fig:ballistic} plots a trajectory of a test particle in the $x_{\rm p}-z_{\rm p}$ plane that begins orbiting the primary star. 
The particle starts  along the line joining the two stars at distance $0.36 a$ from the primary on a circular orbit around the primary that is tilted
by $i_{\rm p0}=60^\circ$ from the binary orbital plane.
 The particle undergoes KL oscillations and reaches a large enough eccentricity to be transferred to the secondary star through Roche lobe overflow. The particle is transferred when it is near apastron and at the part of its orbit that has the highest distance above the binary orbital plane.

To study the behaviour of discs, we perform Smoothed Particle Hydrodynamics (SPH) numerical simulations using the code {\sc phantom} \citep{Lodato2010,Price2010,Price2012,Price2017}.  Misaligned discs in binary systems have been extensively studied with {\sc phantom} (e.g. \cite{Nixon2012,Nixon2013,Nixon2015,Dougan2015}). We perform SPH simulations using $N=5 \times 10^5$ particles. 
The resolution of the simulation depends on $N$, the viscosity parameter $\alpha$ and the disc scale height $H$. 
The \cite{SS1973} viscosity parameter is modelled by adapting artificial viscosity according to the approach of \cite{Lodato2010}. 
The main implication of this treatment is that in order to provide a constant $\alpha$ in the disc, the disc scale height $H$ must be uniformly resolved.
This is achieved by choosing both a surface density profile and a temperature profile that ensures this condition \citep{Lodato2007}, as discussed below. 

Disc evolution strongly depends on the value of the protoplanetary discs viscosity which is not well known.
In our simulations, we model disc viscosity using the artificial viscosity parameter $\alpha_{\rm AV}$ \citep{Lodato2010}.
The main consequence of modelling the disc viscosity through the artificial viscosity parameter is that there is a lower limit below which a physical viscosity is not resolved in SPH ($\alpha_{\rm AV}\approx 0.1$). Viscosities smaller than this value can produce disc spreading that is independent of the value of $\alpha_{\rm AV}$ \citep{Bate1995,Meru2012}.

Previous simulations of discs undergoing KL oscillations were performed with a locally isothermal temperature profile centered on the primary star. Since we are interested in disc formation around the secondary,  we use a temperature profile that ensures that the locally isothermal temperature distribution is valid close to either star. The circumprimary and circumsecondary disc temperatures should be dominated by the primary and secondary star irradiation, respectively. We adopt the
sound speed distribution (proportional to the square root of the temperature) of \cite{Farris2014} and
 set the sound speed $c_{\rm s}$  to be 
\begin{equation}
c_{\rm s} =  c_{\rm s0} \left( \frac{a}{M_1+M_2}\right)^{q} \left(\frac{M_1}{R_1} + \frac{M_2}{R_2}\right)^{q},
\label{eq:farris}
\end{equation}
 where $R_{\rm i} = |\bm{R} - \bm{R_{\rm i}}|$ with ${\rm i} = 1,2$ are the radial distances from the primary and the secondary star, respectively, and $c_{\rm s0}$ 
is a constant of proportionality.
The sound speed given by Equation (\ref{eq:farris}) ensures that close to primary (secondary) the temperature of the disc is set by the primary (secondary), while for $R_{i} \gg a$ the sound speed is set by the distance from the centre of mass of the binary.
Furthermore, the distribution ensures a smooth transition for material that moves from one star to the other.
For $q=0$ the sound speed is constant and the disc is globally isothermal. For a disc of vertical scale height $H$ with $q=0.5$, the disc aspect ratio $H/R_{i}$ is constant in radius close to either
star, $R_i \ll a$.
With the exception of the simulations in Section \ref{sec:temp}, we take $q = 0.75$.
We consider an initial configuration with a highly misaligned accretion disc around only one component of a circular equal mass binary.  The binary has total mass $M=M_1 + M_2=1$ and a circular orbit in the $x$-$y$ plane with separation $a$.
We choose the accretion radius for the particle removal to be $R_{\rm acc}=0.025a$ around each component of the binary. The mass and angular momentum of any particle that enters the sink radius are added to that of the sink particle. 

The circumprimary disc initially has a mass of $M_{\rm d}=0.001\,M$ and it is inclined with respect to the binary orbital plane. The initial surface density profile is set as $\Sigma\propto (R_1/R_{\rm in})^{-3/2}(1-\sqrt{R_{\rm in}/R_1})$ between $R_{\rm in}=0.025a$ and $R_{\rm out}=0.25a$ 

With this density distribution and a sound speed distribution given by Equation (\ref{eq:farris}) for $q=0.75$, the disc is uniformly resolved. The circumsecondary disc is taken to be initially absent, i.e., its density distribution is initially zero.
Tides from the companion star can overcome the viscous torque and truncate the disc \citep{Paczynski1977,Goldreich1980,Papaloizou1977,Artymowicz1994}. 
These torques transfer angular momentum from the disc to the orbit of the binary.
The outer radius is chosen to be the tidal truncation radius for the disc assuming a coplanar binary.
Since misaligned discs feel a weaker torque compared to coplanar discs, the outer edge of the disc might be larger than this value \citep{Lubow2015,Miranda2015,Nixon2015}. 
We vary the other parameters: disc aspect ratio, viscosity, initial inclination, and binary mass ratio (see Table \ref{tab:accmass}).

In the simulations, we identify particles as belonging to the circumprimary, circumsecondary, and circumbinary discs at a given time. To do this, we calculate the specific energy of each particle to see whether it is bound to the central object within the Roche lobe it occupies.
To determine which particles belong to the  circumbinary disc (see Section \ref{sec:circbin}), we require that the particles orbit around the binary and are energetically bound to it. Since both the circumprimary and the newly formed circumsecondary disc respond in a global fashion when subjected to KL oscillations in these simulations, we consider
as representative values the averages of both the eccentricity and inclination over the whole disc radial extent. 

\subsection{Disc formation and effects of the temperature distribution}
\label{sec:temp}

We show that mass transfer can occur from the primary disc to form a secondary disc. We also explore the effects of the disc temperature profile on the mass transfer process. We compare the results of the simulations with the same viscosity coefficient $\alpha=0.1$ and an equal mass binary, but using  different equations of state. The sound speed profiles are described by Equation (\ref{eq:farris}).
We apply a globally isothermal profile ($q=0$, run1) and a spatially varying temperature profile   with $q=0.5$ (run2). The different $q$ values change the temperature structure, that in turn change the disc aspect ratios about each star $i$, $H/R_{i}$, where $H$ is the disc scale height.
The globally isothermal equation of state implies the disc aspect ratio increases with radius, while Equation (\ref{eq:farris}) implies a constant disc aspect ratio for $q = 0.5$. 
For comparison, we choose circumprimary discs that have the same aspect ratio at radius $R_1=R_{\rm out}$. We compare a globally isothermal disc with $H/R_1\,(R_{\rm out})=0.03$ with a locally isothermal disc ($q=0.5$) with constant $H/R_1 =0.03$ (i.e. we compare run1 with run2). In these cases, the disc aspect ratios at the initial inner edges are $H/R_1\,(R_{\rm in})=0.01$ and 0.03 for the $q=0$ and $q=0.5$ cases, respectively.

The bottom panel of Figure \ref{fig:macc_eos_14_1} plots  the total amount of mass accreted into the sink about the secondary (within a distance
of $R_{\rm acc}=0.025a$ of the secondary) in units of the initial circumprimary disc mass, $\Delta M/M_{\rm d}$, as a function of time in units of the binary orbital period.
The amount of mass accreted onto the secondary star is about a factor of 2 higher  in the locally isothermal $q=0.5$ case. This increase is due to the fact that the amplitude of the KL oscillations of the disc eccentricity is larger in the constant disc aspect ratio case ($q=0.5$) than in the globally isothermal  case ($q=0$). 

The four upper panels of Figure \ref{fig:macc_eos_14_1} show the eccentricity and inclination evolution of the circumprimary (left panels) and circumsecondary (right panels) discs. 
In both cases, the circumsecondary disc forms with non-zero initial eccentricity and inclination above the critical KL angle $i_{\rm cr}$ for an initially circular test particle orbit.
We show the eccentricity and inclination of the circumsecondary disc only for the $q=0.5$  locally isothermal simulation because of the lack of resolution for the circumsecondary disc in the globally isothermal case.
In the $q=0.5$ simulation (black line) we see that the circumsecondary disc undergoes a KL oscillation. 

The KL oscillation period of the circumsecondary disc in the $q=0.5$ case is  shorter than the circumprimary disc oscillation period, in agreement with the test particle calculations presented in Section \ref{sec:eccKL}.
 The minimum tilt angle of the circumsecondary disc is significantly below the critical angle $i_{\rm cr} \simeq 39^\circ$ for initially circular test particle orbits.
Minimum tilt angles below $i_{\rm cr}$ can be produced for test particles that start with a nonzero eccentricity, as discussed in Section \ref{sec:eccKL}.
 But since the circumprimary disc inclination also reaches below
$i_{\rm cr}$, other effects such as disc dissipation or gas pressure may play a role. 

Figure \ref{fig:mdisc_eos14_1} plots the amount of mass contained in the circumprimary (upper panel) and circumsecondary (lower panel) discs in units of the initial circumprimary disc mass $M_{\rm d}=0.001$ as a function of time for the simulation with sound speed given by Equation (\ref{eq:farris}) with $q=0.5$ (run2). Notice that, while the lowest panel of Figure~\ref{fig:macc_eos_14_1} plots $\Delta M/M_{\rm d}$, the amount of mass accreted within a distance $R_{\rm acc}$ from the secondary star,  Figure \ref{fig:mdisc_eos14_1} plots the amount of mass within each Roche-lobe outside distance $R_{\rm acc}$ that is also energetically bound to each star. 
The circumsecondary disc starts with zero mass and it acquires mass as the first KL oscillation begins. After roughly $22\,P_{\rm b}$ the circumsecondary disc has been almost completely accreted onto the secondary star.  


In the remainder of this work we use the \cite{Farris2014} sound speed distribution given by Equation (\ref{eq:farris}) with $q=0.75$ to describe the temperature profile of the circumprimary disc and to model the circumsecondary disc formed as a consequence of the transfer of mass through Roche lobe overflow.    

\subsection{Effects of the circumprimary disc aspect ratio}
\label{sec:hoverr}

The amount of mass transferred  to the secondary star depends crucially on  the value of the disc aspect ratio of the circumprimary disc. The main effects that play a role are essentially two: the disc has to spread fast enough in order for its outer edge to fill the Roche lobe and at the same time the material in the circumprimary disc cannot be drained too quickly through accretion onto the primary star. The disc aspect ratio affects both processes. A lower (higher) value of $H/R_1$ corresponds to slower (faster) viscous spreading, while also leading to a lower (higher) accretion rate onto the primary star.

\begin{figure}
\centering
	\includegraphics[width=\columnwidth]{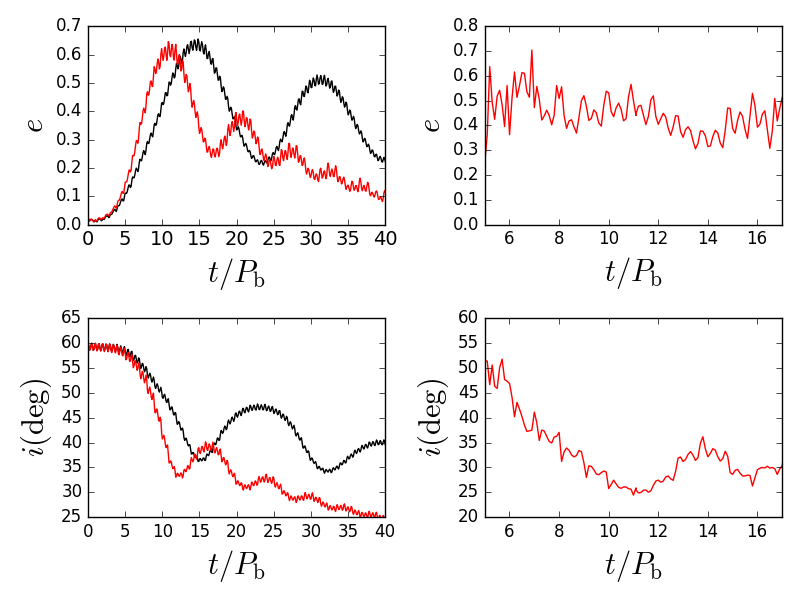}
    \includegraphics[width=0.8\columnwidth]{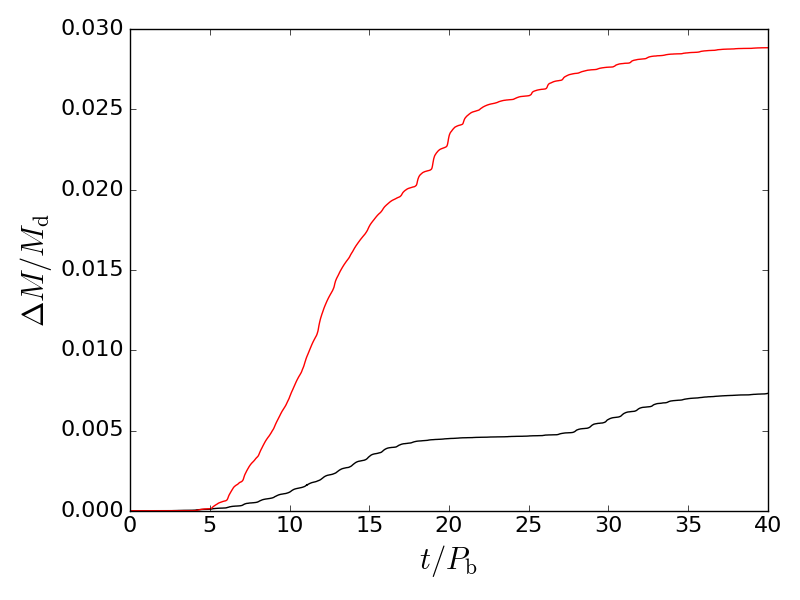}
    \caption{Top four panels: Eccentricity (upper panels) and inclination (lower panels) evolution of the circumprimary (left) and circumsecondary (right) disc averaged over the whole disc. 
    Bottom panel: Accreted mass within $R_{\rm acc}=0.025a$ of the secondary star scaled by the initial circumprimary disc mass $\Delta M / M_{\rm d}$ as a function of time in units of $P_{\rm b}$. 
    The black curves refer to a circumprimary aspect ratio $H/R_1(R_{\rm in})=0.01$ (run3) while the red lines are for $H/R_1(R_{\rm in})=0.035$ (run4).  The sound speed varies in radius according to  Equation (\ref{eq:farris}) with $q=0.75$.}
    \label{fig:hoverr}
\end{figure}

\begin{figure}
\centering
\includegraphics[width=0.8\columnwidth]{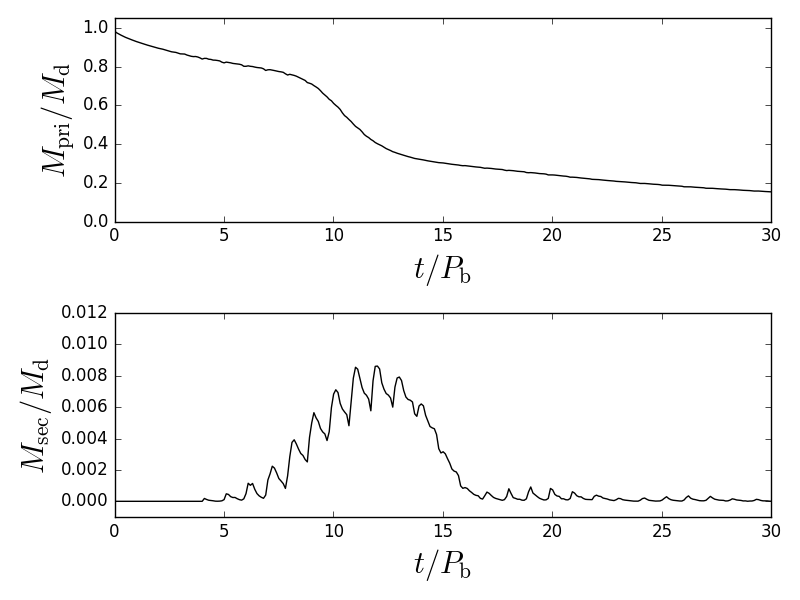}
\caption{Mass contained in the circumprimary (upper panel) and circumsecondary (lower panel) discs outside of $R_{\rm acc}$ of the central stars in units of the initial circumprimary disc mass $M_{\rm d}=0.001$ as a function of time in units of $P_{\rm b}$. The disc sounds speed is described in Equation (\ref{eq:farris}) with $q=0.75$ and $H/R_1(R_{\rm in})=0.035$  (run4).}
\label{fig:mdisc_hoverr}
\end{figure}

High disc sound speeds can suppress KL oscillations.
\cite{Zanazzi2017} and \cite{LO2017} showed that the critical inclination angle for KL oscillations to occur changes with the disc aspect ratio. According to  Fig. 2 of \cite{LO2017}, if the disc aspect ratio $H/R \ga 0.15$  at the disc outer edge in an equal mass binary, then there are typically no KL modes present, due to the effects of pressure on the disc apsidal precession. Thus, the disc would not undergo KL oscillations and there would be no mass transfer between the components of the binary. 

Figure \ref{fig:hoverr} shows the effect of changing the disc aspect ratio with other parameters held constant in simulations and $q=0.75$ in Equation (\ref{eq:farris}). We consider disc aspect ratios at the disc inner edge  $H/R_1(R_{\rm in}) = 0.01$ (run3) and $H/R_1(R_{\rm in}) = 0.035$ (run4), represented by the black and red curve, respectively. The top panels show the eccentricity (upper panel) and the inclination (lower panel) of the primary (left) and secondary (right) discs. 
The smaller disc aspect ratio (black curves) leads to higher amplitude KL oscillations in the primary disc. The oscillations are slightly delayed and have a longer period. This delay may explain in part the smaller amount of mass being transferred to the secondary star during the $40\,P_{\rm b}$ of the simulation. In both cases, the disc aspect ratios $H/R_1$ at the disc inner edges are small enough to prevent fast accretion onto the primary star.  
In addition, the disc aspect ratios at the disc outer edges are also small enough to prevent fast outward spreading of the disc. The higher $H/R_1=0.035$ case results in faster outward spreading than in the lower  $H/R_1=0.01$ which may also explain the higher mass accretion rate onto the secondary for the higher $H/R_1$ case.

As seen in Figures~\ref{fig:hoverr}  and \ref{fig:mdisc_hoverr}, the transfer of mass between the two binary components is triggered after roughly $5\,P_{\rm b}$, when the circumprimary disc eccentricity begins to grow. 
In the left upper panels of Figure \ref{fig:hoverr} we show the eccentricity and inclination of the circumsecondary disc only for the higher aspect ratio simulation (run4). Again, we see evidence for tilt oscillations with circumsecondary disc periods that are shorter
that circumprimary disc periods, as discussed in Section \ref{sec:temp}.
For the lower aspect ratio simulation (run3), there are not enough particles around the secondary to properly resolve the accretion disc behaviour and so we do not plot the corresponding results for that case.

As a test of the importance of KL oscillations to the mass transfer process, we ran a model with an initial tilt that is too low for KL oscillations to occur. 
We compare the amount of mass accreted onto the secondary star in the higher aspect ratio simulation (run4) that has an initial disc inclination of $60^\circ$ to that obtained with the same parameters, but with the initial circumprimary disc inclination of $40^{\circ}$. In this case, the disc does not undergo KL oscillations and the amount of mass accreted onto the companion is negligibly small ($\Delta M/M_{\rm d} \sim 10^{-5}$) is  after $40\,P_{\rm b}$.
Therefore, it is clear that the mass transfer mechanism is KL oscillations.




\begin{table*}
	\centering
    \caption{Parameters of the circumprimary disc for each simulation. The first column contains the label for each simulation, the second column is the slope of the circumprimary disc temperature profile, the third column contains the disc aspect ratio at the inner edge, the fourth column is the binary mass ratio, the fifth and sixth columns contain the disc viscosity and initial inclination respectively. Column~7 shows the binary eccentricity and column~8 shows the circumprimary disc outer radius is units of the binary separation.  Column~9 shows the amount of mass accreted onto the secondary star after a time of $40\,P_{\rm b}$, $\Delta M/M_{\rm d}$. The last column contains the respective figures in the paper for each simulation.}
	\begin{tabular}{lcccccccll} 
	\hline
ID & $q$ & $H/R_1(R_{\rm in})$ & $M_2/M_1$ & $\alpha$ & $i$ & $e_{\rm b}$ & $R_{\rm out}/a$ & $\Delta M/M_{\rm d}$ & Fig. \\
	\hline
 	run1 & $0$ & $0.01$ & $1.0$ & $0.1$ & $60^{\circ}$ & 0 & $0.25$ & $0.018$& \ref{fig:macc_eos_14_1}\\
   	run2 & $0.5$ & $0.03$ & $1.0$ & $0.1$ & $60^{\circ}$ & 0 & $0.25$ & $0.038$& \ref{fig:macc_eos_14_1}, \ref{fig:mdisc_eos14_1}  \\
   	run3 & $0.75$ & $0.01$ & $1.0$ & $0.1$ & $60^{\circ}$ & 0 & $0.25$ & $0.0078$& \ref{fig:hoverr}\\
    run4 & $0.75$ & $0.035$ & $1.0$ & $0.1$ & $60^{\circ}$ & 0 & $0.25$ & $0.029$& \ref{fig:hoverr}, \ref{fig:mdisc_hoverr}, \ref{fig:macc_alpha}, \ref{fig:massratio}\\
    run5 & $0.75$ & $0.035$ & $1.0$ & $0.01$ & $60^{\circ}$ & 0 & $0.25$ & $0.0065$&\ref{fig:macc_alpha}\\
    run6 & $0.75$ & $0.035$ & $0.5$ & $0.1$ & $60^{\circ}$ & 0 & $0.25$ & $0.042$&\ref{fig:massratio}, \ref{fig:mdisc_massratio}\\
   	run7 & $0.75$ & $0.05$ & $1.0$ & $0.1$ & $70^{\circ}$ & 0 & $0.25$ & $0.11$& \ref{fig:7c},\ref{fig:par_incl}, \ref{fig:mdisc_incl}\\
    run8 & $0.75$ & $0.035$ & $1.0$ & $0.1$ & $70^{\circ}$ & 0 & $0.25$ & $0.055$& \\
    run9 & $0.75$ & $0.05$ & $1.0$ & $0.1$ & $70^{\circ}$ & 0.5 & $0.2$ & $0.03$&  \ref{fig:ebin}\\
	\hline
	\end{tabular}
    \label{tab:accmass}
\end{table*}

\begin{figure}
\centering
	\includegraphics[width=0.8\columnwidth]{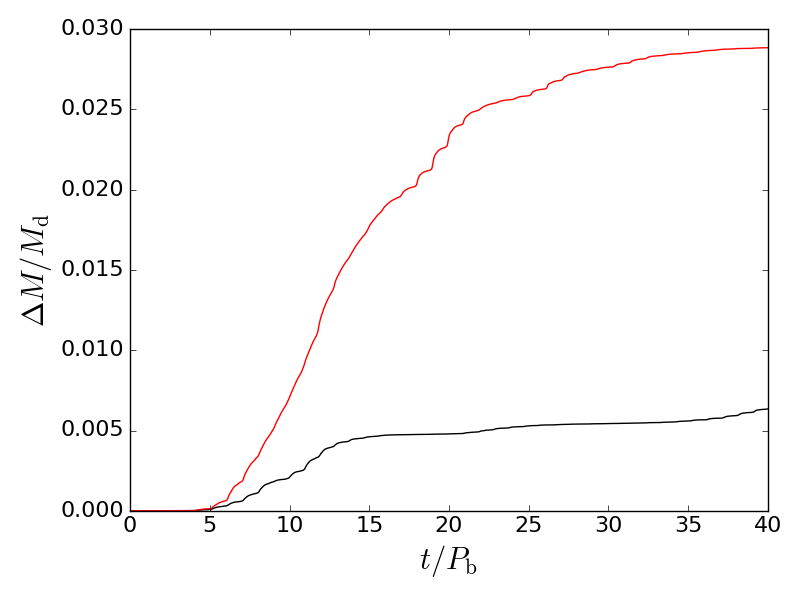}
    \caption{Accreted mass within a radius $R_{\rm acc}$ of the secondary star scaled by the initial circumprimary disc mass $\Delta M / M_{\rm d}$ as a function of time in units of $P_{\rm b}$. The black line refers to $\alpha=0.01$ (run5) while the red line refers to $\alpha=0.1$ (run4). The sound speed varies in radius according to  Equation (\ref{eq:farris}) with $q=0.75$.}
    \label{fig:macc_alpha}
\end{figure}
%

\begin{figure}
\centering
	\includegraphics[width=\columnwidth]{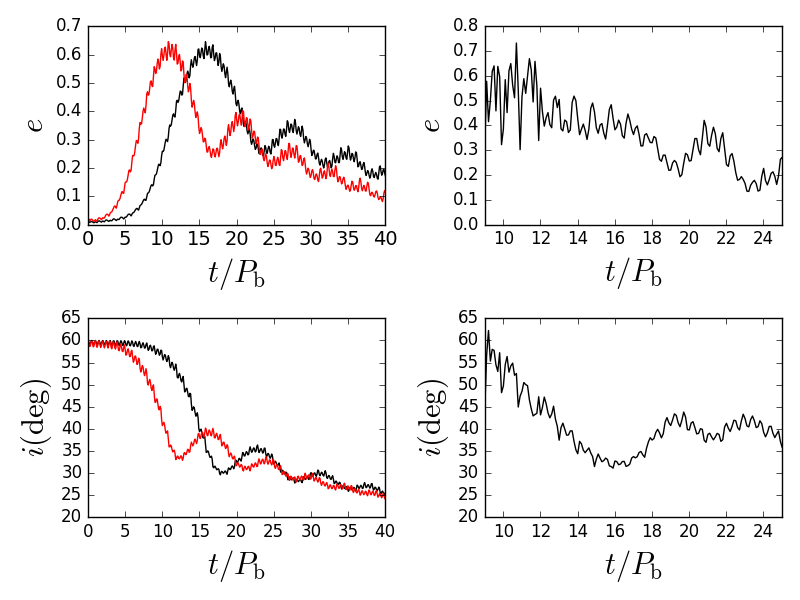}
    \includegraphics[width=0.8\columnwidth]{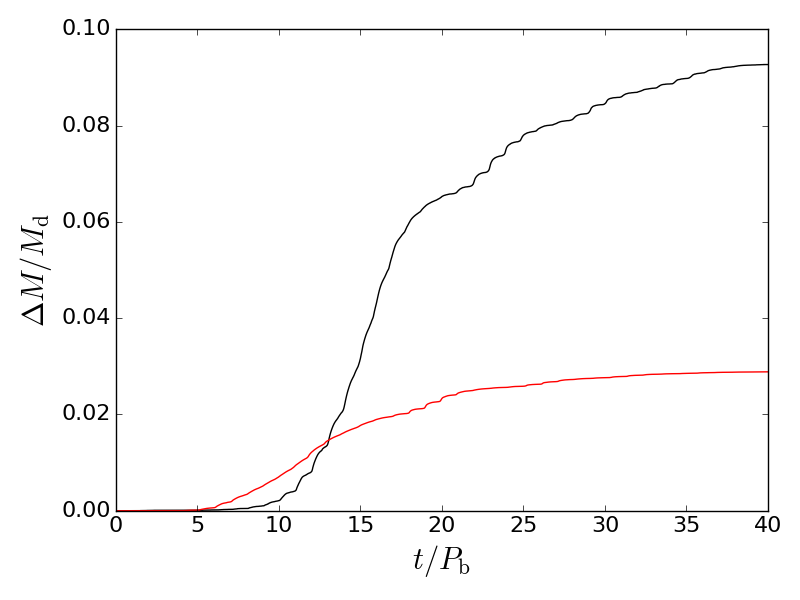}
    \caption{Top four panels: Eccentricity (upper panels) and inclination (lower panels) evolution of the circumprimary (left) and circumsecondary (right) disc as a function of time in units of $P_{\rm b}$ averaged over the whole disc. 
    Bottom panel: Accreted mass within a distance of $R_{\rm acc}=0.025a$ from the secondary star scaled by the initial circumprimary disc mass $\Delta M / M_{\rm d}$ as a function of time in units of $P_{\rm b}$. The black and red lines refer to mass ratio $0.5$ (run6) and $1.0$ (run4) respectively.
    The sound speed varies in radius according to  Equation (\ref{eq:farris}) with $q=0.75$. }
    \label{fig:massratio}
\end{figure} 

\begin{figure}
\centering
\includegraphics[width=0.8\columnwidth]{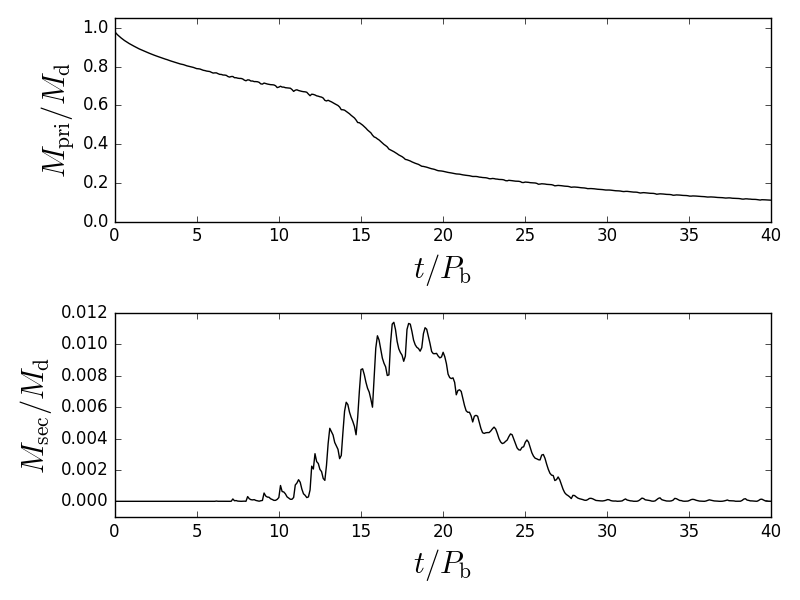}
\caption{Mass contained in the circumprimary (upper panel) and circumsecondary (lower panel)  discs outside of $R_{\rm acc}$ of the central stars in units of the initial circumprimary disc mass $M_{\rm d}=0.001$ as a function of time in units of $P_{\rm b}$ for the simulation in run6.}
\label{fig:mdisc_massratio}
\end{figure}

\subsection{Disc viscosity}
\label{sec:visc}

The disc viscosity provides the source of angular momentum transport causing the disc to spread outwards on the viscous timescale $t_{\nu} = R^2/\nu$. According to the $\alpha-$prescription \citep{SS1973},  for a fixed gas temperture distribution a lower value of $\alpha$ implies slower disc spreading. 
Figure \ref{fig:macc_alpha} shows the amount of accreted mass within radius $R_{\rm acc}$ of the secondary normalised by  the initial circumprimary disc mass $\Delta M/M_{\rm d}$ as a function of time, in units of the binary period, for two different values of the viscosity parameter $\alpha=0.01$ and $0.1$ (i.e. we compare run4 and run5), represented by the black and red curves respectively.
Decreasing the viscosity by an order of magnitude results in a decrease in the amount of mass accreted onto the secondary by roughly a factor of $3$. With a smaller viscosity, the circumprimary disc spreads out more slowly. Since the timescale for the period of  the KL oscillations decreases with distance from the primary, the smaller primary disc has a longer KL oscillation period.  Furthermore, the smaller disc leads to less mass transfer to the secondary.
However, the mass transfer onto the secondary star is not suppressed, but is delayed in time. A smaller viscosity value leads also to slower accretion onto the primary star. Therefore, there would be more material in the circumprimary disc that can eventually be transferred to the secondary star.

Note that the values of the viscosity parameter used in this work, $\alpha=0.01$ and 0.1, are large compared to the value inferred from the observations of protoplanetary discs, i.e. $10^{-4}-10^{-3}$ \citep[e.g.,][]{Pinte2016, Flaherty2017}. As we discussed in the introduction, the inability to model such low viscosity values with SPH codes is a limitation. However, the dynamical effect of viscosity depends on the kinematic viscosity $\nu$ which depends on the disc aspect ratio, as well as $\alpha$. While we cannot set $\alpha$ to a very small value, we can lower the disc aspect ratio to obtain the same correspondent kinematic viscosity. For instance, in the simulation run5 we obtain the same kinematic viscosity of a typical protostellar disc with $H/R=0.1$ and $\alpha \sim 10^{-4}$.

\subsection{Binary mass ratio}
\label{sec:mu}

We consider the effects of changing the binary mass ratio on the mass 
transfer process.
In the bottom panel of  Fig.~\ref{fig:massratio} we compare the accreted mass onto the secondary star for two different values of the mass ratio $\mu=0.5$ and $1.0$ (i.e. we compare run6 and run4). 
The figure shows that the amount of mass transferred to the secondary increases for the small mass ratio case. This effect may be due to the weaker torque exerted by a less massive companion that acts to prevent gas from entering its Roche lobe. But other effects may play a role, such as the properties of the KL oscillations. 

The upper four panels of  Figure \ref{fig:massratio}  show the eccentricity and inclination of the circumprimary (left panels) and circumsecondary (right panels) disc evolution with time.  A smaller mass ratio leads to an increased timescale for the KL oscillations. The disc is eccentric for longer so it can transfer more mass onto the secondary star (see bottom panel in Figure \ref{fig:massratio}).
We plot the eccentricity and inclination evolution of the circumsecondary disc of the simulation with $\mu=0.5$ (run6), which is to be compared with the equal mass binary simulation (run4) in Figure \ref{fig:hoverr}. 

As also seen in the upper panels of Figure \ref{fig:massratio}, the effect of decreasing the mass ratio of the binary components $\mu =M_2/M_1$ is essentially to cause a delay in the KL oscillations and therefore a delay in the mass transfer, as was also found by \cite{Fu2015a}.  
 
In  the simulations, this delay may be explained by the fact that the initial disc around the primary has to spread farther out for the torque exerted by the secondary to be strong enough to trigger KL oscillations. 

Furthermore, we see a hint of KL oscillations of the circumsecondary disc in the simulation with binary mass ratio $\mu = 0.5$ at around $20\,P_{\rm b}$. This  has a period of about $5\,P_{\rm b}$ which is roughly half the KL oscillation period of the (initially circular) circumprimary disc.
Figure \ref{fig:mdisc_massratio} shows the amount of mass contained in the circumprimary (upper panel) and circumsecondary (lower panel) disc in units of the initial circumprimary disc mass $M_{\rm d}=0.001$ as a function of time for the simulation with lower binary mass ratio (run6). 


\begin{figure*}
 \includegraphics[width=0.9\textwidth]{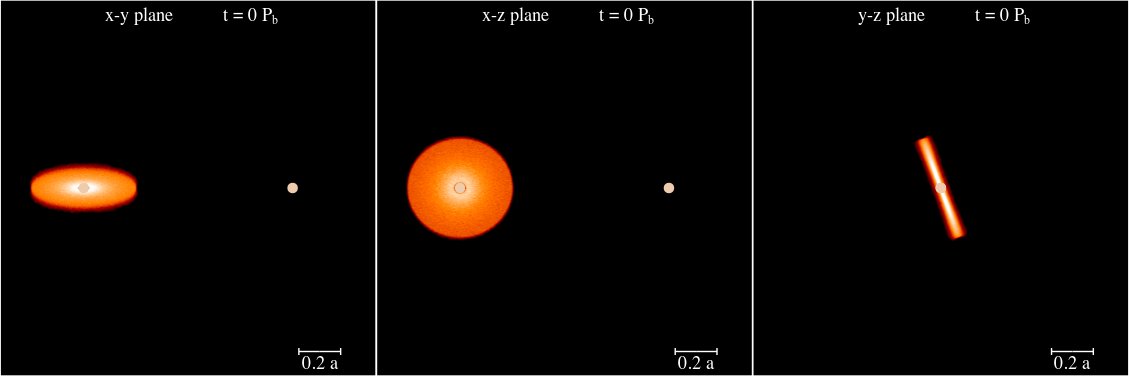}
	\includegraphics[width=0.9\textwidth]{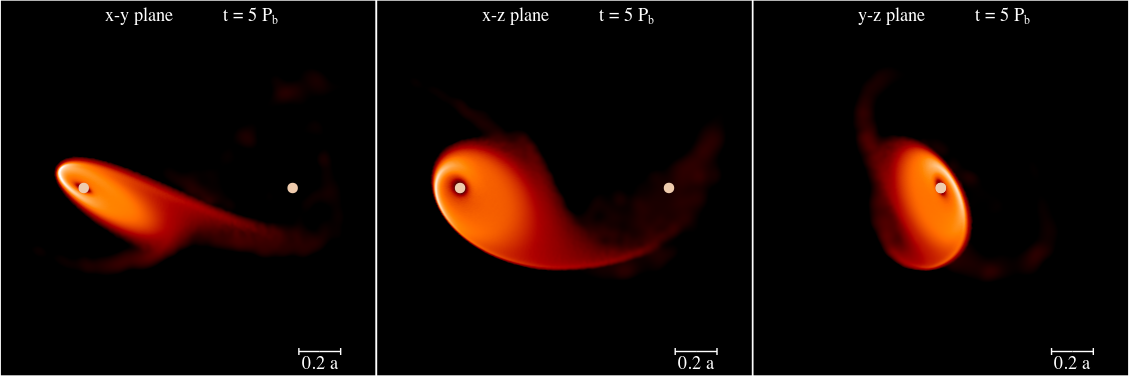}
    \includegraphics[width=0.9\textwidth]{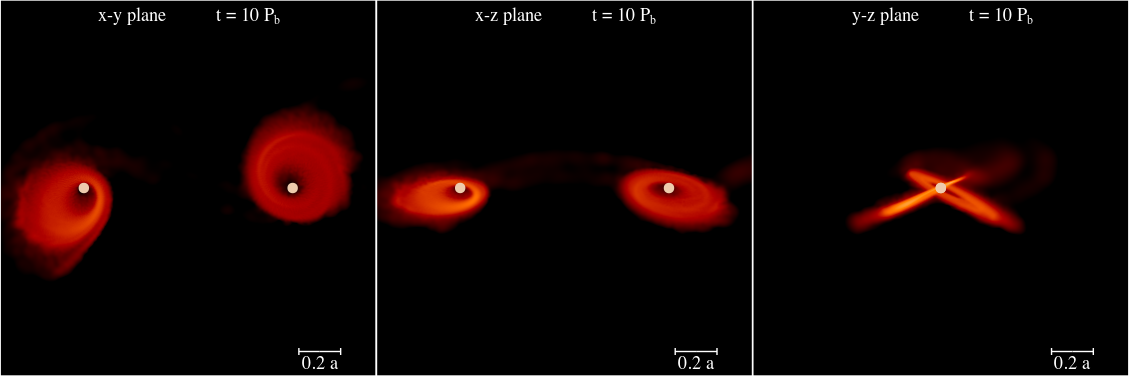}
    \includegraphics[width=0.9\textwidth]{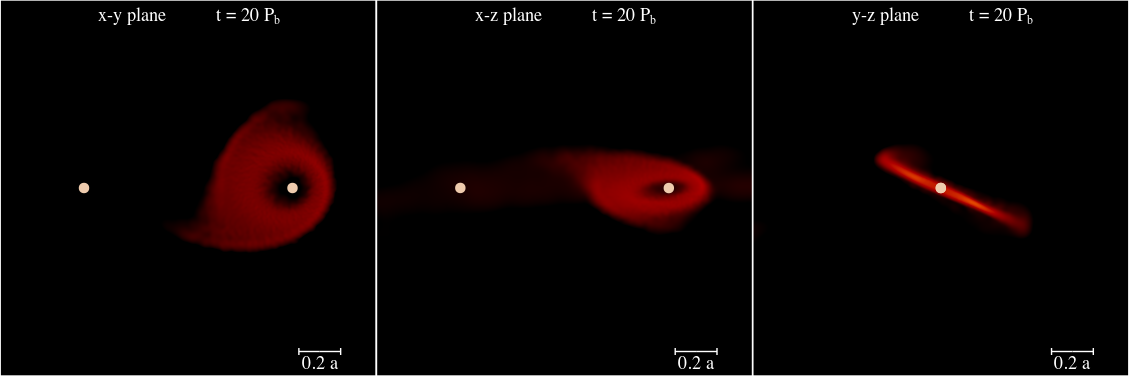}
    \caption{  SPH simulation of a binary (shown by the white circles) with a disc around the primary star forming a circumsecondary disc through KL oscillations. The system is plotted in the frame that is corotating with the binary. The size of the circles denotes the accretion radius of the SPH sink particle. The panels show the view looking down the $x-y$, $x-z$ and $y-z$ planes. The binary orbit lies in the $x-y$ plane. The circumprimary is initially tilted from the binary orbital plane by 70$^{\circ}$ and becomes eccentric due to the KL mechanism. The parameters are the fiducial ones with sound speed defined by Equation (\ref{eq:farris}) and $H/R_1(R_{\rm in})=0.05$. The panels refer to times $t =5$ (upper),$10$ (middle) and $20\, P_{\rm b}$ (lower). Only one star is shown in the right hand panels because the stars lie on top of each other.}
    \label{fig:7c}
\end{figure*}

\begin{figure}
\centering
	\includegraphics[width=\columnwidth]{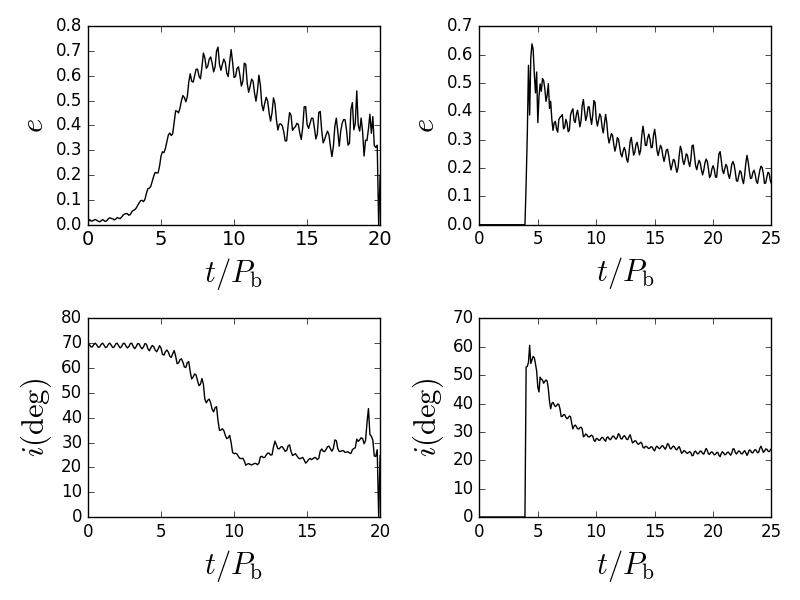}
    \includegraphics[width=0.8\columnwidth]{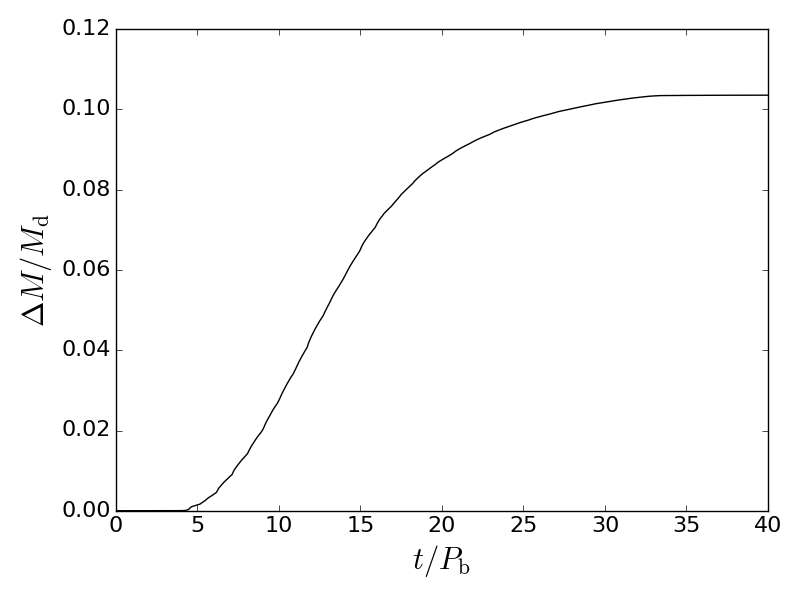}
    \caption{Top four panels: Eccentricity (upper panels) and inclination (lower panels) evolution of the circumprimary (left) and circumsecondary (right) disc as a function of time in units of $P_{\rm b}$ averaged over the whole disc for a disc with an initial inclination of $70^\circ$ (run7).
    Bottom panel: Accreted mass within a distance of $R_{\rm acc}=0.025a$ from the secondary star scaled by the initial circumprimary disc mass $\Delta M / M_{\rm d}$ as a function of time in units of $P_{\rm b}$. }
    \label{fig:par_incl}
\end{figure}

\begin{figure}
\centering
\includegraphics[width=0.8\columnwidth]{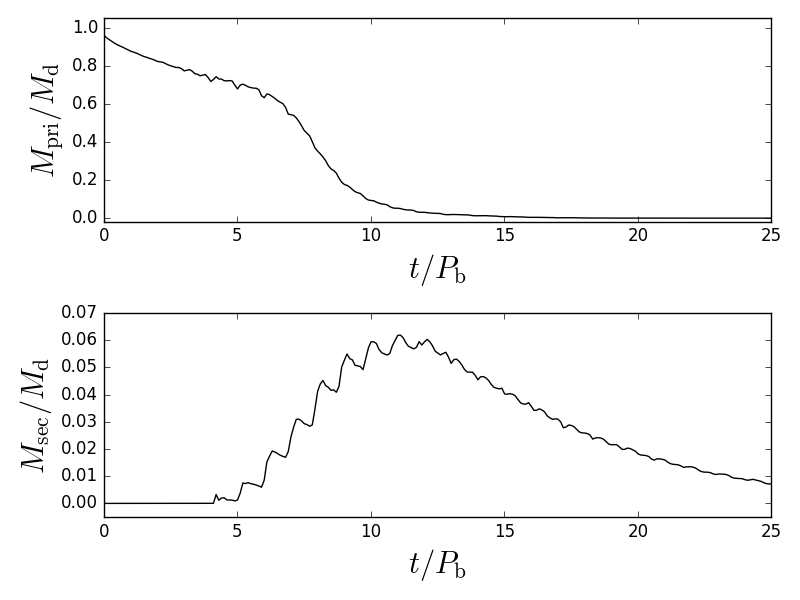}
\caption{Mass contained in the circumprimary (upper panel) and circumsecondary (lower panel)  discs outside of $R_{\rm acc}$ of the central stars in units of the initial circumprimary disc mass $M_{\rm d}=0.001$ as a function of time in units of $P_{\rm b}$ for the simulation in run7.}
\label{fig:mdisc_incl}
\end{figure}

\subsection{Circumprimary disc initial misalignment angle}
\label{sec:incl}

We investigate the effect of the initial inclination angle of the circumprimary accretion disc on the amount of mass that is transferred to the secondary star and how it affects the circumsecondary disc formation.
 Using the similar model parameters to run4 in Table 1 and a higher disc aspect ratio $H/R_1(R_{\rm in})=0.05$,  we increase the initial inclination of the disc around the primary to $i=70^{\circ}$ for run7.

We first compare the amount of mass transferred for different simulations with the same aspect ratio and different circumprimary disc initial inclination to isolate the effect of the latter. We found that a higher inclination leads to an increase of the mass transferred by roughly a factor 2 (comparison between run7 and run8 in Table \ref{tab:accmass}). Since we now know that a higher disc aspect ratio at the inner edge leads to higher mass transfer, we ran the same simulation with a higher value of $H/R_1(R_{\rm in})$ to better resolve the circumsecondary disc.

We define a Cartesian coordinate system whose origin is at the binary centre of mass and corotates with the binary. The $x$-axis is along the direction from the center of mass to the secondary star.  The $z-$axis is along the direction of the binary angular momentum. 
The results of the SPH simulation run7 are shown in Figure~\ref{fig:7c}. The surface density of the accretion discs is shown in the $x-y$, $x-z$, and $y-z$ planes for times of $t =5,10,$ and $20\, P_{\rm b}$. 

Figure~\ref{fig:7c} shows how the disc formation process operates.
The KL oscillations of the circumprimary disc result in an increase of its eccentricity from its initially circular form.
The apastron distance increases in the outer disc to be large enough for some material to enter into the Roche lobe of the companion. It can clearly be seen that the material is transferred from the primary to the secondary.
The flow occurs in the form of a gas stream.
The stream of material that flows into the secondary Roche lobe has angular momentum about the secondary and therefore does not directly impact it. This material is on an eccentric orbit around the secondary. The stream of material self-intersects and dissipates energy via shocks. 
 Both discs can be seen to be highly eccentric and misaligned at $t=10\,P_{\rm b}$. At a time $t = 20\,P_{\rm b}$, only the secondary disc  contains enough mass to be visible on the plot.
Since the circumprimary disc is inclined with respect to the binary orbit, the material enters the Roche lobe of the  secondary with a non-zero inclination and therefore the resulting accretion disc is misaligned to the binary orbital plane. 

The bottom panels of Figure \ref{fig:par_incl} and Figure \ref{fig:mdisc_incl} show that there is more mass transferred to the secondary than in the previous cases that have an initial inclination of $60^\circ$.
As seen in the upper left panel, increasing the  initial disc inclination leads to more accretion within the accretion radius of the primary star because the disc becomes more eccentric during the KL oscillations (see Equation (\ref{eq:emax})). 
The circumprimary disc is completely accreted after roughly $20 \,P_{\rm b}$ which corresponds to two KL oscillations of its eccentricity and inclination. However, the amplitude of the oscillations is larger. Thus, the mass transfer to the secondary is more efficient.
As seen in the upper right panel of
Figure \ref{fig:par_incl}, the circumsecondary disc starts forming after $5 \,P_{\rm b}$ when the first KL oscillation occurs. 
The amount of mass transferred to the secondary star after $40\,P_{\rm b}$ is $\Delta M/M_{\rm d} = 0.11$. 

The comparison between the upper four left and right panels of Figure \ref{fig:par_incl} shows that an initially eccentric accretion disc experiences KL oscillations on a shorter period compared to the circular case, as discussed in Section \ref{sec:temp}. 

Figure \ref{fig:mdisc_incl} shows the amount of mass contained in the circumprimary (upper panel) and circumsecondary (lower panel) disc in units of the initial circumprimary disc mass $M_{\rm d}=0.001$ as a function of time. The circumsecondary disc starts with zero mass and it acquires mass as the first KL oscillation begins. The circumsecondary disc contains more mass than the circumprimary disc at  times later than $\sim 12\,P_{\rm b}$. After roughly $25\,P_{\rm b}$ the circumsecondary disc has been largely accreted onto the secondary star.

\begin{figure}
\centering
	\includegraphics[width=0.8\columnwidth]{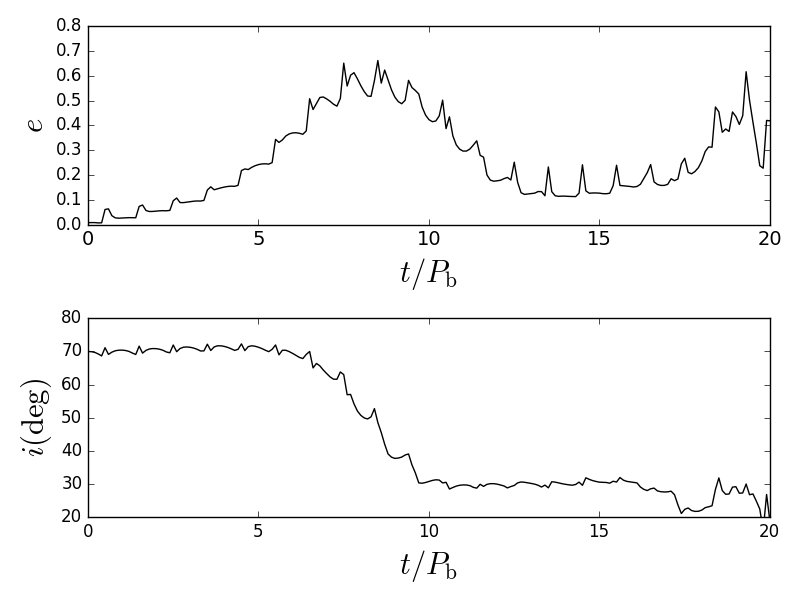}
    \caption{Eccentricity (upper panel) and inclination (lower panel) evolution of the circumprimary  disc as a function of time in units of $P_{\rm b}$ averaged over the whole disc for a disc with an initial inclination of $70^\circ$, disc outer radius $R_{\rm out}=0.2a$, and binary eccentricity $e_{\rm b}=0.5$ (run9).}
    \label{fig:ebin}
\end{figure}

\subsection{Binary eccentricity}
\label{sec:ebin}

The observed binary orbital eccentricities vary with binary orbital period \citep{Raghavan2010, Tokovinin2016}.  The closest binaries with periods less than 10 days  
are circularised by tidal dissipation. The average binary eccentricity increases as a function of binary orbital period and ranges from about $0.4$ to $0.6$.
We investigate the effect of the binary eccentricity on the mass transfer from the circumprimary disc by running a simulation (run9) with the same parameters as run7 of Section \ref{sec:incl} that involves a circular orbit binary,  except that we take the binary eccentricity to be  $e_{\rm b}=0.5$. We also reduced the circumprimary disc outer radius to $R_{\rm out}=0.2a$, since the torque exerted by the eccentric binary is stronger and this results in a smaller truncation radius \citep{Artymowicz1994, Fu2015a}.

Figure \ref{fig:ebin} shows the circumprimary disc eccentricity and inclination evolution as a function of time in units of the binary orbital period. After $20\,P_{\rm b}$ the circumprimary disc is accreted.
The amount of mass transferred to the secondary star, roughly 3\%, is much lower than we find for a circular binary in run7, 11\%. We do not show the evolution of the newly formed circumsecondary disc, since we do not have enough resolution to see KL oscillations. However, as in the circular case, the circumsecondary disc forms with non-zero eccentricity ($e\approx0.6$) and at a high inclination ($i\approx 60^{\circ}$). So it is likely that this disc is subject to KL oscillations.

\section{Circumbinary disc formation}
\label{sec:circbin}

\begin{figure}
\centering
	\includegraphics[width=0.8\columnwidth]{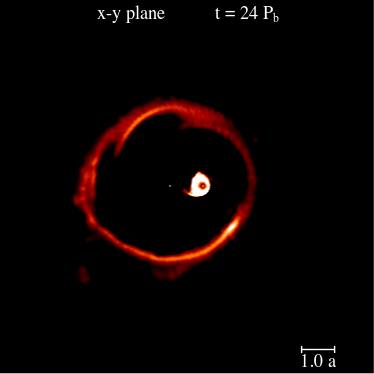}
    \caption{SPH simulation of a binary (shown by the red circles) with a disc around the primary star forming a circumsecondary and a circumbinary disc through KL oscillations (run7). The size of the red circles denotes the accretion radius of the SPH sink particles. The surface density is plotted in the $x-y$ plane after $24\,P_{\rm b}$. }
    \label{fig:7c_bin}
\end{figure}

\begin{figure}
\centering
	\includegraphics[width=0.8\columnwidth]{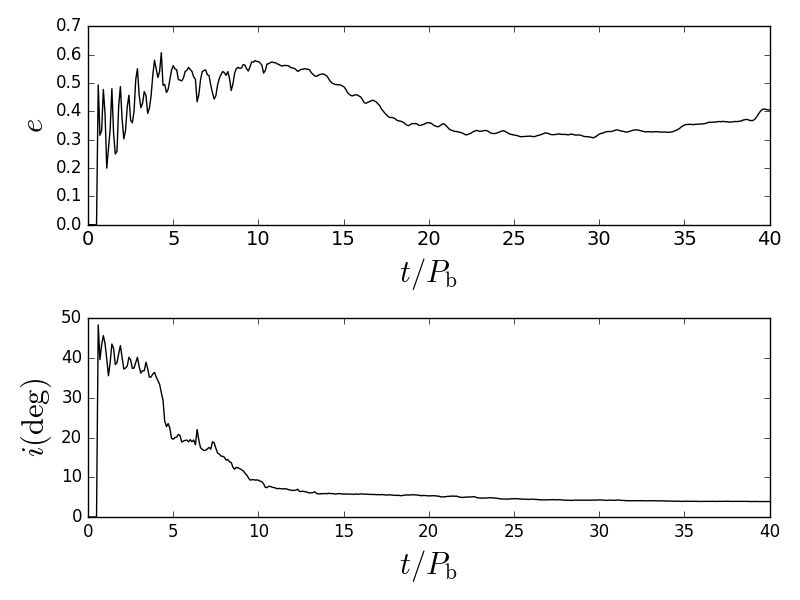}
    \includegraphics[width=0.8\columnwidth]{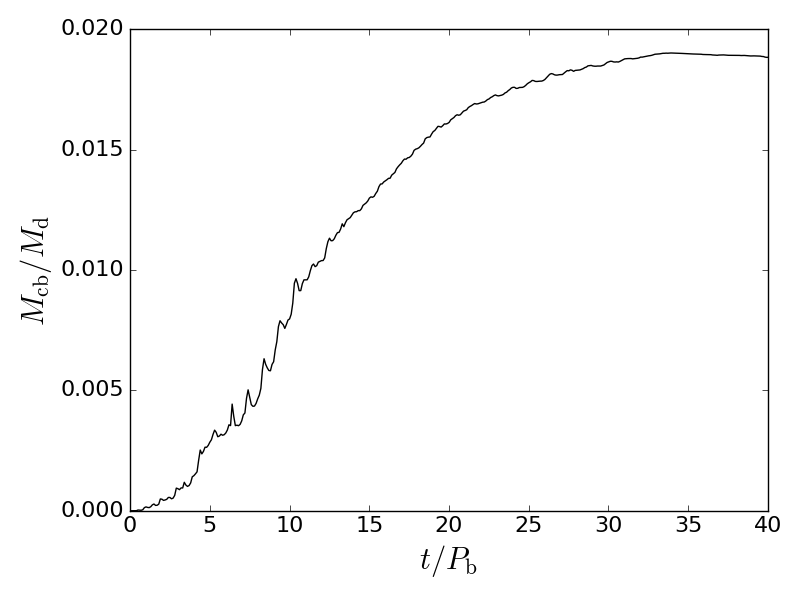}
    \caption{Upper and middle panel: Eccentricity and inclination evolution of the circumbinary disc. 
    Bottom panel: Amount of mass contained in the circumbinary disc as a function of time, in units of the initial circumprimary disc mass.
    The parameters of the circumprimary disc are  $\alpha=0.1,$ and $H/R(R_{\rm in})=0.05$, initial disc tilt $i = 70^{\circ}$, and the binary is equal mass (run7).}
    \label{fig:par_bin}
\end{figure}

As the circumprimary disc undergoes KL oscillations of eccentricity, some of the material that escapes the Roche lobe of the primary is captured in orbit around the binary. 
The same process might also occur though overflow of a pre-existing circumsecondary disc, if it were included in the simulations.
The material self-intersects and circularizes forming a circumbinary disc.
Figure \ref{fig:7c_bin} shows the surface density profile of the circumsecondary and circumbinary disc after $24\,P_{\rm b}$. This profile is the result of the simulation described in Section \ref{sec:incl}. The newly formed circumbinary disc extends from $R_{\rm in}\sim a$ up to roughly $R_{\rm out} \sim 5a$. 
The eccentricity and inclination of the circumbinary disc are shown in Figure \ref{fig:par_bin}, where we average the quantities over the whole disc.

In the simulation we identify particles as belonging to the circumbinary disc if they lie outside the binary and are energetically  bound to the binary.
The circumbinary disc that forms is highly eccentric and mildly inclined with respect to the binary plane. The eccentricity decreases slightly with time while the disc gradually aligns with the binary plane over  $40\,P_{\rm b}$. 
The amount of mass that remains in the circumbinary disc at the end of the simulation is around 1-2\% of the initial circumprimary disc mass if the binary is circular. The lower panel of Figure \ref{fig:par_bin} shows the mass of the circumbinary disc, in units of the circumprimary disc initial mass, as a function of time in units of the binary orbit.

A problem related to planet formation in circumbinary accretion discs in tight binaries is that the discs have been found to be rather compact and might be not massive enough to form giant planets \citep{Harris2012}.
We found the newly formed circumbinary disc to be radially very narrow in our simulations.
The viscous evolution of a compact disc is faster compared to one of an extended disc. Therefore accretion of this disc material onto the binary occurs on a shorter timescale and this could significantly reduce the available time in which planets could form.
Furthermore circumbinary discs formed in this way contain only 1\% of the initial circumprimary disc mass and this could be another obstacle for planet formation. However, it is likely that the material that forms the circumbinary disc in our simulation will add to an already present circumbinary disc. Thus, the mechanism described here increases the amount of circumbinary material.
If the binary has eccentricity of $e_{\rm b}=0.5$ (run9), almost 10\% of the initial circumprimary disc mass is captured by the binary. 
Therefore, there is more material that can add to an already present disc than in the circular binary case.

The newly formed circumbinary disc is likely to spread initially, but  would eventually be accreted onto the binary, as is known for coplanar circumbinary discs \citep{Artymowicz1996}. Furthermore, since this disc forms eccentric and inclined with respect to the orbital plane of the binary, its interaction with an already present disc might lead to a change in its eccentricity and inclination.

\section{Conclusions}
\label{sec:concl}

We analysed the evolution of an initially misaligned disc around the primary star in a circular and eccentric orbit binary system with no initial circumsecondary disc.
We have found that when the circumprimary disc undergoes KL oscillations of eccentricity and inclination, it transfers mass to the secondary that results in the formation of a misaligned circumsecondary disc. 
The circumsecondary disc also undergoes KL oscillations. The two discs are  misaligned with respect to each other as well as with respect to the binary orbital plane. In addition, some circumprimary gas flows outward to form a circumbinary disc.

The amount of mass that is transferred from the primary to the secondary star through KL oscillations depends on several parameters of both the circumprimary disc and the binary itself.
Within the range of parameters we examine, increasing the disc aspect ratio, its viscosity, and  initial inclination angle leads to larger amount of mass transfer onto the secondary. It is generally of order of $\la 10\%$ of the initial primary disc mass.
Decreasing the binary mass ratio also leads to a larger amount of mass transfer.  However, we expect that the behaviour changes at more extreme values
of some of these parameters. For example, at sufficiently high viscosity and small secondary mass, the accretion onto the primary will occur before a KL disc oscillation is well underway. Also, the KL disc mechanism is suppressed at higher disc sound
speeds \citep{Zanazzi2017,LO2017}. 
In such cases, KL mass transfer mechanism will not operate. 
Binary eccentricity appears to decrease the amount of gas captured by the secondary and increase the amount of gas in the circumbinary disc. Further
analysis of the effects of binary eccentricity would be worth pursuing.


The SPH simulations performed in this work are scale free. Therefore, the results could in principle be applied to both wide and close binaries.
The typical size of observed circumstellar discs is up to a few hundred au. Therefore, in the case of HK~Tau, with binary separation $\sim 350\,$au, the KL oscillation mechanism may be operating.
However, if the binary separation is large compared to the disc size it might be difficult for the circumprimary disc to fill the Roche-lobe of the primary star and to create a misaligned disc around the companion star within the lifetime of the primary disc. 	 

It is likely that both components of a binary star system form with an accretion disc and these discs may be misaligned to each other and to the binary orbital plane. In a more complete model,  there will be mass transfer between the two components. The material transferred to each component will change the angular momentum and eccentricity of a disc that already exists. In a future publication, we will investigate the interaction of the accreted material with an already present circumsecondary disc.


\section*{Acknowledgements}

We thank Mario Livio for useful discussions.
We thank Daniel Price for providing the {\sc phantom} code for SPH simulations and acknowledge the use of {\sc splash} \citep{Price2007} for the rendering of the figures. We acknowledge support from NASA through grant NNX17AB96G.  Computer support was provided by UNLV's National Supercomputing Center.




\bibliographystyle{mnras}
\bibliography{mnras1} 


\bsp	
\label{lastpage}
\end{document}